\documentclass{ws-procs9x6}

\begin{document}

\title{Quiver Gauge Theory and Unification at About 4 TeV}

\author{P.H. FRAMPTON}

\address{Department of Physics and Astronomy, \\
University of North Carolina at Chapel Hill, \\ 
Chapel Hill, NC 27599-3255, USA.\\ 
E-mail: frampton@physics.unc.edu}


\maketitle

\abstracts{The use of the AdS/CFT correspondence to arrive at quiver
gauge field theories is dicussed, focusing on the orbifolded
case without supersymmetry. An abelian orbifold with the finite
group $Z_{p}$ can give rise to a $G = SU(N)^p$ gauge group
with chiral fermions and complex scalars in different
bi-fundamental representations of $G$. The precision
measurements at the $Z$ resonance suggest the values
$p = 12$ and $N = 3$, and a unifications scale
$M_U \sim 4$ TeV.The robustness and predictivity of such
grand unification is discussed.}

\section{Quiver Gauge Theory}

The relationship of the Type IIB superstring to conformal gauge theory
in $d=4$ gives rise to an interesting class of gauge
theories.
Choosing the simplest compactification\cite{Maldacena}
on $AdS_5 \times S_5$ gives rise to an $N = 4$ SU(N) gauge theory
which is known to be conformal due to
the extended global supersymmetry and non-renormalization theorems. All
of the RGE $\beta-$functions for this $N = 4$
case are vanishing in perturbation theory. It is possible to break
the $N=4$ to $N=2,1,0$ by replacing
$S_5$ by an orbifold $S_5/\Gamma$
where $\Gamma$ is a discrete group with
$\Gamma \subset SU(2), \subset SU(3), \not\subset SU(3)$
respectively.

In building a conformal gauge theory model \cite{Frampton,FS,FV},
the steps are: (1) Choose the discrete group $\Gamma$; (2) Embed
$\Gamma \subset SU(4)$; (3) Choose the $N$ of $SU(N)$; and
(4) Embed the Standard Model $SU(3) \times SU(2) \times U(1)$
in the resultant gauge group $\bigotimes SU(N)^p$ (quiver
node identification). Here we shall look only
at abelian $\Gamma = Z_p$ and define
$\alpha = exp(2 \pi i/p)$. It is expected from the string-field
duality that the resultant field
theory is conformal in the $N\longrightarrow \infty$ limit,
and will have a fixed manifold, or at least a fixed point, for $N$ finite.

Before focusing on $N=0$ non-supersymmetric cases, let
us first examine an $N=1$ model first
put forward in the work of
Kachru and Silverstein\cite{KS}.
The choice is $\Gamma = Z_3$ and the {\bf 4} of $SU(4)$
is {\bf 4} = $(1, \alpha, \alpha, \alpha^2)$. Choosing N=3
this leads to the three chiral families under $SU(3)^3$
trinification\cite{DGG}
\begin{equation}
(3, \bar{3}, 1) + (1, 3, \bar{3}) + (\bar{3}, 1, 3)
\end{equation}
In this model it is interesting that
the number of families arises as 4-1=3,
the difference between the 4 of SU(4) and $N=1$,
the number of unbroken supersymmetries.
However this model has no gauge coupling unification;
also, keeping $N=1$ supersymmetry is against the
spirit of the conformality approach. We now
present an example which accommodates
three chiral families, break all supersymmetries
($N=0$) and possess gauge coupling unification, including the
correct value of the electroweak mixing angle.

Choose $\Gamma = Z_7$, embed the 4 of SU(4)
as $(\alpha^2, \alpha^2, \alpha^{-3}, \alpha^{-1})$, and
choose N=3 to aim at a trinification
$SU(3)_C \times SU(3)_W \times SU(3)_H$.

The seven nodes of the quiver diagram will
be identified as C-H-W-H-H-H-W.

The behavior of the 4 of SU(4) implies that the bifundamentals
of chiral fermions are in the representations
\begin{equation}
\sum_{j=1}^{7} [ 2(N_j, \bar{N}_{j+2}) + (N_j, \bar{N}_{j-3})
+ (N_j, \bar{N}_{j-1})]
\end{equation}
Embedding the C, W and H SU(3) gauge groups as
indicated by the quiver mode identifications
then gives the seven quartets of irreducible representations
\begin{equation}
\begin{array}{l}
[3(3,\bar{3}, 1) + (3, 1, \bar{3})]_1 + \\

+ [3(1, 1, 1+8) + (\bar{3}, 1, 3)]_2 + \\

+ [3(1, 3, \bar{3}) + (1, 1 + 8, 1)]_3 + \\

+ [(2(1, 1, 1+8) + (1, \bar{3}, 3) + (\bar{3}, 1, 3)]_4 + \\

+ [2(1, 1, 1 + 8) + 2 (1, \bar{3}, 3)]_5 + \\

+ [2(\bar{3}, 1, 3) + (1, 1, 1 + 8) + (1, \bar{3}, 3)]_6 + \\

+ [4(1, 3, \bar{3})]_7
\end{array}
\end{equation}
Combining terms gives, aside from (real) adjoints and overall singlets
\begin{equation}
3(3, \bar{3}, 1) + 4(\bar{3}, 1, 3) + (3, 1, \bar{3})
+ 7(1, 3, \bar{3}) + 4(1, \bar{3}, 3)
\end{equation}
Cancelling the real parts (which acquire Dirac masses at the conformal
symmetry breaking scale) leaves under trinification
$SU(3)_C \times SU(3)_W \times SU(3)_H$
\begin{equation}
3 [(3, \bar{3}, 1) + (1, 3, \bar{3}) + (\bar{3}, 1, 3)]
\end{equation}
which are the desired three chiral families.

Given the embedding of $\Gamma$ in SU(4) 
it follows that the 6 of SU(4) transforms as
$(\alpha^4, \alpha, \alpha, \alpha^{-1}, \alpha^{-1}, \alpha^{-4})$.
The complex scalars therefore transform as
\begin{equation}
\sum_{j=1}^{7} [(N_j, \bar{N}_{j\pm4}) + 2 (N_j, \bar{N}_{j\pm1})]
\end{equation}
These bifundamentals can by their VEVS break the symmetry $SU(3)^7$
= $SU(3)_C \times SU(3)_W^2 \times SU(3)^4_H$ down
to the appropriate diagonal subgroup
$SU(3)_C \times SU(3)_W \times SU(3)_H$.

Now to the final aspect of this model 
which is its motivation, the gauge coupling
unification. The embedding in $SU(3)^7$ of
$SU(3)_C \times SU(3)^2_W \times SU(3)^4_H$ means that the
couplings $\alpha_1, \alpha_2, \alpha_3$ are
in the ratio
$\alpha_1 / \alpha_2 / \alpha_3 = 1/2/4$.
Using the phenomenological data given at the beginning,
this implies that
$sin^2\theta = 0.231$. On the
other hand, the QCD coupling is
$\alpha_3 = 0.0676$ which is too low unless the
conformal scale is at least 10TeV.

\section{Gauge Couplings.}

An alternative to conformality, grand unification with supersymmetry,
leads to an impressively accurate gauge coupling unification\cite{ADFFL}.
In particular it predicts an electroweak mixing angle
at the Z-pole, ${\tt sin}^2 \theta = 0.231$. This result
may, however, be fortuitous, but rather than
abandon gauge coupling unification, we can rederive ${\tt sin}^2 \theta = 0.231$
in a different way by embedding the electroweak $SU(2) \times U(1)$ in
$SU(N) \times SU(N) \times SU(N)$ to
find ${\tt sin}^2 \theta = 3/13 \simeq 0.231$\cite{FV,F2}.
This will be a common feature of the models in this paper.

The conformal theories will be finite without
quadratic or logarithmic divergences. This requires appropriate
equal number of fermions and bosons which can cancel in
loops
and which occur without the necessity of space-time supersymmetry.
As we shall see in one example, it is possible to combine
spacetime supersymmetry
with conformality but the latter is the driving principle and the former
is merely an option: additional fermions and scalars are predicted by
conformality in the TeV range\cite{FV,F2},
but in general these particles are different and distinguishable
from supersymmetric partners.
The boson-fermion cancellation is essential for the
cancellation of infinities, and will
play a central role in the calculation of the cosmological constant
(not discussed here). In the field picture, the
cosmological constant measures the vacuum energy density.

What is needed first for the conformal approach is a simple
model.

Here we shall focus on abelian orbifolds characterised by the discrete group
$Z_p$. Non-abelian orbifolds will be systematically analysed elsewhere.

The steps in building a model for the abelian case (parallel steps
hold for non-abelian orbifolds) are:

\begin{itemize}

\item{(1)} Choose the discrete group $\Gamma$. Here we are considering
only $\Gamma = Z_p$. We define $\alpha = {\rm exp}(2 \pi i/p)$.

\item{(2)} Choose the embedding of $\Gamma \subset SU(4)$ by
assigning ${\bf 4} = (\alpha^{A_1}, \alpha^{A_2}, \alpha^{A_3}, \alpha^{A_4})$
such that $\sum_{q=1}^{q=4} A_q = 0 ({\rm mod} p)$. To
break $N = 4$ supersymmetry to
$N = 0$ ( or $N = 1$) requires that
none (or one) of the $A_q$ is equal to zero (mod p).

\item{(3)} For chiral fermions one requires that ${\bf 4} \not\equiv {\bf 4^{*}}$
for the embedding of $\Gamma$ in $SU(4)$.

The chiral fermions are in the bifundamental representations of $SU(N)^p$
\begin{equation}
\sum_{i=1}^{i=p} \sum_{q=1}^{q=4}
(N_i, \bar{N}_{i + A_q})
\label{fermions}
\end{equation}
If $A_q=0$ we interpret $(N_i, \bar{N}_i)$ as a singlet plus an adjoint
of $SU(N)_i$.

\item{(4)} The {\bf 6} of $SU(4)$ is real {\bf 6} = $(a_1, a_2, a_3, -a_1, -a_2, -a_3)$
with
$a_1 = A_1 + A_2$,
$a_2 = A_2 + A_3$,
$a_3 = A_3 + A_1$
(recall that all components are defined modulo p).
The complex scalars are in the bifundamentals
\begin{equation}
\sum_{i=1}^{i=p} \sum_{j=1}^{j=3}
(N_i, \bar{N}_{i \pm a_j})
\label{scalars}
\end{equation}
\noindent The condition in terms of $a_j$ for $N = 0$ is
$\sum_{j=1}^{j=3} (\pm a_j) \not= 0 ({\rm mod}~~ p)$\cite{Frampton}.
\item{(5)} Choose the $N$ of $\bigotimes_i SU(Nd_i)$ (where the $d_i$ are the
dimensions of the representrations of $\Gamma$). For the abelian case
where $d_i \equiv 1$, it is natural to choose $N=3$ the largest
$SU(N)$ of the standard model (SM) gauge group. For a non-abelian $\Gamma$
with $d_i \not\equiv 1$ the choice $N=2$ would be indicated.

\item{(6)}  The $p$ quiver nodes are identified as color (C), weak isospin (W)
or a third $SU(3)$ (H). This specifies the embedding of the gauge group
$SU(3)_C \times SU(3)_W \times SU(3)_H \subset \bigotimes SU(N)^p$.

This quiver node identification is guided by (7), (8) and (9) below.

\item{(7)}  The quiver node identification is required to
give three chiral families under Eq.(\ref{fermions})
It is sufficient to make three of the $(C + A_q)$ to be W and the fourth H, given that there is only
one C quiver node, so that there are three $(3, \bar{3}, 1)$. Provided that
$(\bar{3}, 3, 1)$ is avoided by the $(C - A_q)$ being H, the remainder
of the three family trinification will be automatic by chiral anomaly cancellation.
Actually, a sufficient condition for three families has been given; it is
necessary only that the difference between the number of $(3 + A_q)$
nodes and the number of $(3 - A_q)$ nodes
which are W is equal to three.

\item{(8)}  The complex scalars of Eq. (\ref{scalars}) must be sufficient for their
vacuum expectation values (VEVs) to spontaneously break
$SU(3)^p \longrightarrow SU(3)_C \times SU(3)_W \times SU(3)_H
\longrightarrow SU(3)_C \times SU(2)_W \times U(1)_Y \longrightarrow SU(3)_C \times U(1)_Q$.

Note that, unlike grand unified theories (GUTs) with or without supersymmetry,
the Higgs scalars are here prescribed by the conformality condition.
This is more satisfactory because it implies that the Higgs sector cannot be chosen
arbitrarily, but it does make model building more interesting

\item{(9)} Gauge coupling unification should apply at least to the electroweak mixing
angle ${\rm sin}^2 \theta = g_Y^2 / (g_2^2 + g_Y^2) \simeq 0.231$. For trinification
$Y = 3^{-1/2} ( - \lambda_{8W} + 2\lambda_{8H})$ so that $(3/5)^{1/2} Y$ is
correctly normalized. If we make $g_Y^2 = (3/5)g_1^2$ and $g_2^2 = 2 g_1^2$
then ${\rm sin}^2 \theta = 3/13 \simeq 0.231$ with sufficient accuracy.

\end{itemize}

\bigskip
\bigskip

We now answer all these steps for the choice $\Gamma = Z_p$
for successive $p = 2, 3 ...$ up to $p = 7$.

\bigskip
\bigskip

\begin{itemize}

\item{{\bf p = 2}}

In this case $\alpha = -1$ and therefore one cannot
costruct any complex {\bf 4} of $SU(4)$ with
${\bf 4} \not\equiv {\bf 4^*}$.
Chiral fermions are therefore
impossible.

\item{{\bf p = 3}}

The only possibilities are $A_q = (1, 1, 1, 0)$ or $A_q = (1, 1, -1, -1)$.
The latter is real and leads to no chiral fermions.
The former leaves $N = 1$ supersymmetry and is a simple three-family
model\cite{KS} by the quiver node identification C - W - H. The scalars $a_j = (1, 1, 1)$
are sufficient to spontaneously break to the SM. Gauge coupling unification is, however,
missing since ${\rm sin}^2 \theta = 3/8$, in
bad disagreement with experiment.

\item{{\bf p = 4}}

The only complex $N = 0$ choice is $A_q = (1, 1, 1, 1)$. But
then $a_j = (2, 2, 2)$ and any quiver node identification
such as C - W - H - H has 4 families and the scalars are insufficient to
break spontaneously the symmetry to the SM gauge group.

\item{{\bf p = 5}}

The two inequivalent complex choices are $A_q = (1, 1, 1, 2)$ and $A_q = (1, 3, 3, 3)$.
By drawing the quiver, however, and using the rules
for three chiral families given in (7)
above, one finds that the node identification and the prescription of the scalars
as $a_j = (2, 2, 2)$ and $a_j = (1, 1, 1)$ respectively does not permit
spontaneous breaking to the standard model.

\item{{\bf p = 6}}

Here we can discuss three inequivalent complex possibilities as follows:

\noindent (6A) $A_q = (1, 1, 1, 3)$ which implies $a_j = (2, 2, 2)$.

\bigskip

\noindent Requiring three families means a node identification C - W - X - H - X - H
where X is either W or H. But whatever we choose for the X the scalar representations
are insufficient to break $SU(3)^6$ in the desired fashion down to the SM. This
illustrates the difficulty of model building when the scalars are not
in arbitrary representations.

\noindent (6B) $A_q = (1, 1, 2, 2)$ which implies $a_j = (2, 3, 3)$.

\bigskip

\noindent
Here the family number can be only zero, two or four as can be seen by inspection
of the $A_q$ and the related quiver diagram. So (6B) is of no phenomenological interest.

\noindent (6C) $A_q = (1, 3, 4, 4)$ which implies $a_j = (1, 1, 4)$.

\bigskip

\noindent Requiring three families needs a quiver node identification which is of the form
{\it either}
C - W - H - H - W - H {\it or} C - H - H - W - W - H. The scalar representations
implied by $a_j = (1, 1, 4)$ are, however, easily seen to be insufficient to do the required
spontaneous symmetry breaking (S.S.B.) for both of these identifications.

\item{{\bf p =7}}

Having been stymied mainly by the rigidity of the scalar representation
for all $p \leq 6$, for $p = 7$ there
are the first cases which work. Six inequivalent complex embeddings of $Z_7
\subset SU(4)$ require consideration.

\noindent (7A) $A_q = (1, 1, 1, 4) \Longrightarrow a_j = (2, 2, 2)$

\bigskip

For the required nodes C - W - X - H - H - X - H the
scalars are {\it insufficient} for S.S.B.

\bigskip

\noindent (7B) $A_q = (1, 1, 2, 3) \Longrightarrow a_j = (2, 3, 3)$

\bigskip

The node identification C - W - H - W - H - H - H leads
to a {\it successful} model.

\bigskip

\noindent (7C) $A_q = (1, 2, 2, 2) \Longrightarrow a_j = (3, 3, 3)$

\bigskip

Choosing C - H - W - X - X - H - H to derive three families, the
scalars {\it fail} in S.S.B.

\bigskip

\noindent (7D) $A_q = (1, 3, 5, 5) \Longrightarrow a_j = (1, 1, 3)$

\bigskip

The node choice C - W - H - H - H - W - H leads
to a {\it successful} model. This is Model A of \cite{F2}.

\bigskip

\noindent (7E) $A_q = (1, 4, 4, 5) \Longrightarrow a_j = (1, 2, 2)$

\bigskip

The nodes C - H - H - H - W - W - H are
{\it successful}.

\bigskip

\noindent (7F) $A_q = (2, 4, 4, 4) \Longrightarrow a_j = (1, 1, 1)$

\bigskip

Scalars {\it insufficient} for S.S.B.

\end{itemize}

The three successful models (7B), (7D) and (7E) lead to
an $\alpha_3(M) \simeq 0.07$. Since $\alpha_3(1 {\rm TeV}) \geq 0.10$
it suggests a conformal scale $M \simeq 10$ TeV\cite{F2}.
The above models have less generators than an
$E(6)$ GUT and thus $SU(3)^7$ merits
further study. It is possible, and under investigation, that non-abelian
orbifolds will lead to a simpler model.

\bigskip

For such field theories it is important to establish the existence of a fixed
manifold with respect to the renormalization group. It
could be a fixed line but more likely, in
the $N = 0$ case, a fixed point. It is known that in the
$N \longrightarrow \infty$ limit the theories become conformal, but
although this 't Hooft limit is where the field-string duality
is derived we know that finiteness survives to finite N in the
$N = 4$ case\cite{mandelstam} and this makes it plausible
that at least a conformal point occurs also for the $N = 0$
theories with $N = 3$ derived above.

The conformal structure cannot by itself predict all the dimensionless ratios of
the standard model such as mass ratios and mixing angles
because these receive contributions,
in general, from soft breaking of conformality. With a
specific assumption about the pattern of conformal
symmetry breaking, however, more work should lead to
definite predictions for such quantities.

\section{4 TeV Grand Unification}

Conformal invariance in two dimensions has had
great success in comparison to several condensed matter
systems. It is an
interesting question whether conformal symmetry
can have comparable success in a four-dimensional
description of high-energy physics.

Even before the standard model (SM)
$SU(2) \times U(1)$ electroweak theory
was firmly established by experimental
data, proposals were made
\cite{PS,GG} of models which would subsume it into
a grand unified theory (GUT) including also the dynamics\cite{GQW} of
QCD. Although the prediction of
SU(5) in its minimal form for the proton lifetime
has long ago been excluded, {\it ad hoc} variants thereof
\cite{FG} remain viable.
Low-energy supersymmetry improves the accuracy of
unification of the three 321 couplings\cite{ADF,ADFFL}
and such theories encompass a ``desert'' between the weak
scale $\sim 250$ GeV and the much-higher GUT scale
$\sim 2 \times 10^{16}$ GeV, although minimal supersymmetric
$SU(5)$ is by now ruled out\cite{Murayama}.

Recent developments in string theory are suggestive
of a different strategy for unification of electroweak
theory with QCD. Both the desert and low-energy
supersymmetry are abandoned. Instead, the
standard $SU(3)_C \times SU(2)_L \times U(1)_Y$
gauge group is embedded in a semi-simple gauge
group such as $SU(3)^N$ as suggested by gauge
theories arising from compactification of the IIB superstring
on an orbifold $AdS_5 \times S^5/\Gamma$ where
$\Gamma$ is the abelian finite group $Z_N$\cite{Frampton}.
In such nonsupersymmetric quiver gauge theories
the unification of couplings happens not by
logarithmic evolution\cite{GQW} over an
enormous desert covering, say, a dozen orders
of magnitude in energy scale. Instead the
unification occurs abruptly at $\mu = M$ through the
diagonal embeddings of 321 in $SU(3)^N$\cite{F2}.
The key prediction of such unification shifts from
proton decay to additional particle content,
in the present model
at $\simeq 4$ TeV.

Let me consider first the electroweak group
which in the standard model is still un-unified
as $SU(2) \times U(1)$. In the 331-model\cite{PP,PF}
where this is extended to $SU(3) \times U(1)$
there appears a Landau pole at $M \simeq 4$ TeV
because that is the scale at which ${\rm sin}^2
\theta (\mu)$ slides to the value
${\rm sin}^2 (M) = 1/4$.
It is also the scale at which the custodial gauged
$SU(3)$ is broken in the framework
of \cite{DK}.

Such theories involve only electroweak
unification so to include QCD I examine the running
of all three of the SM couplings with $\mu$ as
explicated in {\it e.g.} \cite{ADFFL}.
Taking the values at the Z-pole
$\alpha_Y(M_Z) = 0.0101, \alpha_2(M_Z) = 0.0338,
\alpha_3(M_Z) = 0.118\pm0.003$ (the errors in
$\alpha_Y(M_Z)$ and $\alpha_2(M_Z)$ are less than 1\%)
they are taken to run between $M_Z$ and $M$ according
to the SM equations
\begin{eqnarray}
\alpha^{-1}_Y(M) & = & (0.01014)^{-1} - (41/12 \pi) {\rm ln} (M/M_Z)
\nonumber \\
& = & 98.619 - 1.0876 y
\label{Yrun}
\end{eqnarray}
\begin{eqnarray}
\alpha^{-1}_2(M) & = & (0.0338)^{-1} + (19/12 \pi) {\rm ln} (M/M_Z)
\nonumber \\
& = & 29.586 + 0.504 y
\label{2run}
\end{eqnarray}
\begin{eqnarray}
\alpha^{-1}_3(M) & = & (0.118)^{-1} + (7/2 \pi) {\rm ln} (M/M_Z)
\nonumber \\
& = & 8.474 + 1.114 y
\label{3run}
\end{eqnarray}
where $y = {\rm log}(M/M_Z)$.

The scale at which ${\rm sin}^2 \theta(M) = \alpha_Y(M)/
(\alpha_2(M) + \alpha_Y(M))$ satisfies ${\rm sin}^2 \theta (M)
= 1/4$ is found from Eqs.(\ref{Yrun},\ref{2run})
to be $M \simeq 4$ TeV as stated in the introduction above.

I now focus on the ratio $R(M) \equiv \alpha_3(M)/\alpha_2(M)$
using Eqs.(\ref{2run},\ref{3run}). I find
that $R(M_Z) \simeq 3.5$ while $R(M_{3}) = 3$,
$R(M_{5/2}) = 5/2$ and
$R(M_2)=2$ correspond to
$M_3, M_{5/2}, M_2 \simeq 400 {\rm GeV}, ~~ 4 {\rm TeV}, {\rm and}~~ 140 {\rm TeV}$
respectively.
The proximity of $M_{5/2}$ and $M$, accurate to a few percent,
suggests strong-electroweak unification at $\simeq 4$ TeV.

There remains the question of embedding such unification
in an $SU(3)^N$ of the type described in \cite{Frampton,F2}.
Since the required embedding of $SU(2)_L \times U(1)_Y$
into an $SU(3)$ necessitates $3\alpha_Y=\alpha_H$
the ratios of couplings at $\simeq 4$ TeV
is: $\alpha_{3C} : \alpha_{3W} :  \alpha_{3H} :: 5 : 2 : 2$
and it is natural
to examine $N=12$ with diagonal embeddings of
Color (C), Weak (W) and Hypercharge (H)
in $SU(3)^2, SU(3)^5, SU(3)^5$ respectively.

To accomplish this I specify the embedding
of $\Gamma = Z_{12}$ in the global $SU(4)$ R-parity
of the $N = 4$ supersymmetry of the underlying theory.
Defining $\alpha = {\rm exp} ( 2\pi i / 12)$ this specification
can be made by ${\bf 4} \equiv (\alpha^{A_1}, \alpha^{A_2},
\alpha^{A_3}, \alpha^{A_4})$ with $\Sigma A_{\mu} = 0 ({\rm mod} 12)$
and all $A_{\mu} \not= 0$ so that all four supersymmetries are
broken from $N = 4$ to $N = 0$.

Having specified $A_{\mu}$ I calculate the content
of complex scalars by investigating in $SU(4)$
the ${\bf 6} \equiv (\alpha^{a_1}, \alpha^{a_2}, \alpha^{a_3},
\alpha^{-a_3}, \alpha^{-a_2},\alpha^{-a_1})$ with
$a_1 = A_1 + A_2, a_2 = A_2 + A_3, a_3 = A_3 + A_1$ where
all quantities are defined (mod 12).

Finally I identify the nodes (as C, W or H)
on the dodecahedral quiver such that the complex scalars
\begin{equation}
\Sigma_{i=1}^{i=3} \Sigma_{\alpha=1}^{\alpha=12}
\left( N_{\alpha}, \bar{N}_{\alpha \pm a_i} \right)
\label{scalars2}
\end{equation}
are adequate to allow the required symmetry breaking to the
$SU(3)^3$ diagonal subgroup, and the chiral fermions
\begin{equation}
\Sigma_{\mu=1}^{\mu=4} \Sigma_{\alpha=1}^{\alpha=12}
\left( N_{\alpha}, \bar{N}_{\alpha + A_{\mu}} \right)
\label{fermions2}
\end{equation}
can accommodate the three generations of quarks and leptons.

It is not trivial to accomplish all of these requirements
so let me demonstrate by an explicit example.

For the embedding I take $A_{\mu} = (1, 2, 3, 6)$ and for
the quiver nodes take the ordering:
\begin{equation}
- C - W - H - C - W^4 - H^4 -
\label{quiver}
\end{equation}
with the two ends of (\ref{quiver}) identified.

The scalars follow from $a_i = (3, 4, 5)$
and the scalars in Eq.(\ref{scalars2})
\begin{equation}
\Sigma_{i=1}^{i=3} \Sigma_{\alpha=1}^{\alpha=12}
\left( 3_{\alpha}, \bar{3}_{\alpha \pm a_i} \right)
\label{modelscalars}
\end{equation}
are sufficient to break to all diagonal subgroups as
\begin{equation}
SU(3)_C \times SU(3)_{W} \times SU(3)_{H}
\label{gaugegroup}
\end{equation}

The fermions follow from $A_{\mu}$ in Eq.(\ref{fermions2}) as
\begin{equation}
\Sigma_{\mu=1}^{\mu=4} \Sigma_{\alpha=1}^{\alpha=12}
\left( 3_{\alpha}, \bar{3}_{\alpha + A_{\mu}} \right)
\label{modelfermions}
\end{equation}
and the particular dodecahedral quiver
in (\ref{quiver}) gives rise  to exactly {\it three}
chiral generations which transform under (\ref{gaugegroup})
as
\begin{equation}
3[ (3, \bar{3}, 1) + (\bar{3}, 1, 3) + (1, 3, \bar{3}) ]
\label{generations}
\end{equation}
I note that anomaly freedom of the underlying superstring
dictates that only the combination
of states in Eq.(\ref{generations})
can survive. Thus, it
is sufficient to examine one of the terms, say
$( 3, \bar{3}, 1)$. By drawing the quiver diagram
indicated by Eq.(\ref{quiver}) with the twelve nodes
on a ``clock-face'' and using
$A_{\mu} = (1, 2, 3, 6)$
in Eq.(\ref{fermions}) I find
five $(3, \bar{3}, 1)$'s and two $(\bar{3}, 3, 1)$'s
implying three chiral families as stated in Eq.(\ref{generations}).

After further symmetry breaking at scale $M$ to
$SU(3)_C \times SU(2)_L \times U(1)_Y$ the
surviving chiral fermions are the quarks and leptons
of the SM. The appearance
of three families depends on both
the identification of modes in (\ref{quiver})
and on the embedding of $\Gamma \subset SU(4)$. The
embedding must simultaneously give adequate
scalars whose VEVs can break the symmetry
spontaneously to (\ref{gaugegroup}).
All of this is achieved successfully by the
choices made.
The three gauge couplings evolve according to
Eqs.(\ref{Yrun},\ref{2run},\ref{3run}) for
$M_Z \leq \mu \leq M$. For $\mu \geq M$ the
(equal) gauge couplings of $SU(3)^{12}$
do not run if, as conjectured in \cite{Frampton,F2}
there is a conformal fixed point at $\mu = M$.

The basis of the conjecture in \cite{Frampton,F2}
is the proposed duality of Maldacena\cite{Maldacena}
which shows that in the $N \rightarrow \infty$
limit $N = 4$ supersymmetric
$SU(N)$gauge  theory, as well as orbifolded versions with
$N = 2,1$ and $0$\cite{bershadsky1,bershadsky2}
become conformally invariant.
It was known long ago
\cite{mandelstam} that the
$N = 4$ theory is
conformally invariant for all finite $N \geq 2$.
This led to the conjecture in \cite{Frampton}
that the $N = 0$
theories might be conformally
invariant, at least in some case(s),
for finite $N$.
It should be emphasized that this
conjecture cannot be checked
purely
within a perturbative framework\cite{FMink}.
I assume that the local $U(1)$'s
which arise in this scenario
and which would lead to $U(N)$
gauge groups are non-dynamical,
as suggested by Witten\cite{Witten},
leaving $SU(N)$'s.

As for experimental tests of such
a TeV GUT, the situation at energies
below 4 TeV is predicted to be the standard model with
a Higgs boson still to be discovered at a mass
predicted by radiative corrections
\cite{PDG} to be below 267 GeV at 99\% confidence level.

There are many particles predicted
at $\simeq 4$ TeV beyond those of
the minimal standard model.
They include
as spin-0 scalars the states of Eq.(\ref{modelscalars}).
and
as spin-1/2 fermions the states
of Eq.(\ref{modelfermions}),
Also predicted are gauge bosons to fill out the gauge groups
of (\ref{gaugegroup}), and in the same energy region
the gauge bosons to fill out all of
$SU(3)^{12}$. All these extra particles are necessitated by
the conformality constraints of \cite{Frampton,F2} to lie
close to the conformal fixed point.

One important issue is whether this proliferation
of states at $\sim 4$ TeV is
compatible with precision
electroweak data in hand. This has
been studied in the related model of
\cite{DK} in a recent article\cite{Csaki}. Those results
are not easily translated to the present
model but it is possible that such an analysis
including limits on flavor-changing neutral currents
could rule out the entire framework.

As alternative to $SU(3)^{12}$
another approach to TeV unification has as its
group at $\sim 4$ TeV $SU(6)^3$ where
one $SU(6)$ breaks diagonally to color
while the other two $SU(6)$'s each break to
$SU(3)_{k=5}$ where level
$k=5$ characterizes irregular embedding\cite{DM}.
The triangular quiver $-C - W - H - $
with ends identified and $A_{\mu} = (\alpha,
\alpha, \alpha, 1)$, $\alpha = {\rm exp} (2 \pi i / 3)$,
preserves $N = 1$ supersymmetry.
I have chosen to describe the $N = 0$
$SU(3)^{12}$ model in the text
mainly because the symmetry breaking
to the standard model is
more transparent.

The TeV unification fits ${\rm sin}^2\theta$
and $\alpha_3$, predicts three families,
and partially resolves the GUT hierarchy.
If such unification holds in Nature there is a
very rich level of physics one order of
magnitude above presently accessible energy.

Is a hierarchy problem resolved in the present theory?
In the
non-gravitational limit $M_{Planck} \rightarrow \infty$
I have, above the weak scale, the new unification
scale $\sim 4$ TeV. Thus, although not totally resolved,
the GUT hierarchy is ameliorated.

\section{Predictivity}

The calculations have been done in the one-loop
approximation to the renormalization group equations
and threshold effects have been ignored.
These corrections are not expected to be large
since the couplings are weak in the entrire energy
range considered. There are possible further corrections
such a non-perturbative effects, and the effects of
large extra dimensions, if any.

In one sense the robustness of this TeV-scale
unification is almost self-evident, in that it follows from the weakness
of the coupling constants in the evolution from $M_Z$ to $M_U$.
That is, in order to define the theory at $M_U$,
one must combine the effects of
threshold corrections ( due to O($\alpha(M_U)$)
mass splittings )
and potential corrections from redefinitions
of the coupling constants and the unification scale.
We can then {\it impose} the coupling constant relations at $M_U$
as renormalization conditions and this is valid
to the extent that higher order corrections do
not destabilize the vacuum state.

We shall approach the comparison with data in two
different but almost equivalent ways. The first
is "bottom-up" where we use as input that the
values of $\alpha_3(\mu)/\alpha_2(\mu)$ and
$\sin^2 \theta (\mu)$ are expected to be $5/2$
and $1/4$ respectively at $\mu = M_U$.

Using the experimental ranges allowed for
$\sin^2 \theta (M_Z) = 0.23113 \pm 0.00015$,
$\alpha_3 (M_Z) = 0.1172 \pm 0.0020$ and
$\alpha_{em}^{-1} (M_Z) = 127.934 \pm 0.027$
\cite{PDG} we have calculated \cite{FRT}
the values of $\sin^2 \theta (M_U)$
and $\alpha_3 (M_U) / \alpha_2(M_U)$
for a range of $M_U$ between 1.5 TeV
and 8 TeV.
Allowing a maximum discrepancy of $\pm 1\%$ in
$\sin^2 \theta (M_U)$ and
$\pm 4\%$ in $\alpha_3 (M_U) / \alpha_2 (M_U)$
as reasonable estimates of corrections, we deduce that
the unification scale $M_U$ can lie anywhere
between 2.5 TeV and 5 TeV. Thus the theory is
robust in the sense that there is no singular
limit involved in choosing a particular value
of $M_U$.

\bigskip

\noindent Another test of predictivity of the same
model is to fix the unification values at $M_U$ of
$\sin^2 \theta(M_U) = 1/4$ and $\alpha_3 (M_U) /
\alpha_2 (M_U) = 5/2$. We then compute the
resultant predictions at the scale $\mu = M_Z$.

The results are shown for $\sin^2 \theta (M_Z)$
in Fig. 1 with the allowed range\cite{PDG}
$\alpha_3 (M_Z) = 0.1172 \pm 0.0020$. The precise
data on $\sin^2 (M_Z)$ are indicated in Fig. 1 and
the conclusion is that the model makes correct
predictions for $\sin^2 \theta (M_Z)$.
Similarly, in Fig 2, there is a plot of the
prediction for $\alpha_3 (M_Z)$ versus
$M_U$ with $\sin^2 \theta(M_Z)$ held
with the allowed empirical range.
The two quantities plotted in Figs 1 and 2 
are consistent for similar ranges of $M_U$.
Both $\sin^2 \theta(M_Z)$ and $\alpha_3(M_Z)$ are within the empirical limits
if $M_U = 3.8 \pm 0.4$ TeV.

\bigskip

\noindent The model has many additional gauge bosons
at the unification scale, including neutral $Z^{'}$'s,
which could mediate flavor-changing processes
on which there are strong empirical upper limits.

A detailed analysis wll require specific identification
of the light families and quark flavors with
the chiral fermions appearing in the quiver diagram
for the model. We can make only the general
observation that the lower bound on a $Z^{'}$
which couples like the standard $Z$ boson
is quoted as $M(Z^{'}) < 1.5$ TeV \cite{PDG}
which is safely below the $M_U$ values considered here
and which we identify with the mass of the new gauge bosons.

This is encouraging to believe that flavor-changing
processes are under control in the model but
this issue will require more careful analysis when
a specific identification of the quark states
is attempted.

\bigskip

\noindent Since there are many new states predicted
at the unification scale $\sim 4$ TeV, there is a danger
of being ruled out by precision low energy data.
This issue is conveniently studied in terms
of the parameters $S$ and $T$ introduced in \cite{Peskin}
and designed to measure departure from the predictions
of the standard model.

Concerning $T$, if the new $SU(2)$ doublets are
mass-degenerate and hence do not violate a custodial
$SU(2)$ symmetry they contribute nothing to $T$.
This therefore provides a constraint on the spectrum
of new states.

According to \cite{PDG}, a multiplet of degenerate
heavy chiral fermions gives a contribution to $S$:

\begin{equation}
S = C \sum_i \left( t_{3L}(i) - t_{3R}(i) \right)^2 / 3 \pi
\label{capitalS}
\end{equation}
where $t_{3L,R}$ is the third component of weak isopspin
of the left- and right- handed component of
fermion $i$ and $C$ is the number of colors.
In the present model, the additional fermions are non-chiral
and fall into vector-like multiplets and so do not
contribute
to Eq.(\ref{capitalS}).

Provided that the extra isospin multiplets
at the unification scale $M_U$ are sufficiently
mass-degenerate, therefore, there is no conflict
with precision data at low energy.

\section{Discussion}

\bigskip

\noindent The plots we have presented clarify the accuracy
of the predictions of this TeV unification scheme for
the precision values accurately measured at the Z-pole.
The predictivity is as accurate for $\sin^2 \theta$ as
it is for supersymmetric GUT models\cite{ADFFL,ADF,DRW,DG}.
There is, in addition, an accurate prediction for $\alpha_3$
which is used merely as input in SusyGUT models.

At the same time, the accurate predictions are seen to be robust
under varying the unification scale around $\sim 4 TeV$
from about 2.5 TeV to 5 TeV.

In conclusion, since this model ameliorates the GUT hierarchy
problem and naturally accommodates three families, it
provides a viable alternative to the widely-studied
GUT models which unify by logarithmic evolution
of couplings up to much higher GUT scales.

\bigskip
\bigskip

\section*{Acknowledgements}

This work was supported in part by the
Office of High Energy, US Department
of Energy under Grant No. DE-FG02-97ER41036.

\bigskip
\bigskip

\noindent \underline{\bf Figure Captions}

\bigskip

\noindent Fig. 1.

\noindent Plot of $\sin^2 \theta(M_Z)$ versus $M_U$ in TeV, assuming
$\sin^2 \theta(M_U) = 1/4$ and $\alpha_3 / \alpha_2 (M_U) = 5/2$.

\bigskip

\noindent Fig.2.

\noindent Plot of $\alpha_3 (M_Z)$ versus $M_U$ in TeV, assuming
$\sin^2 \theta(M_U) = 1/4$ and $\alpha_3 / \alpha_2 (M_U) = 5/2$.

\newpage

\begin{figure}

\begin{center}

\epsfxsize=4.0in
\ \epsfbox{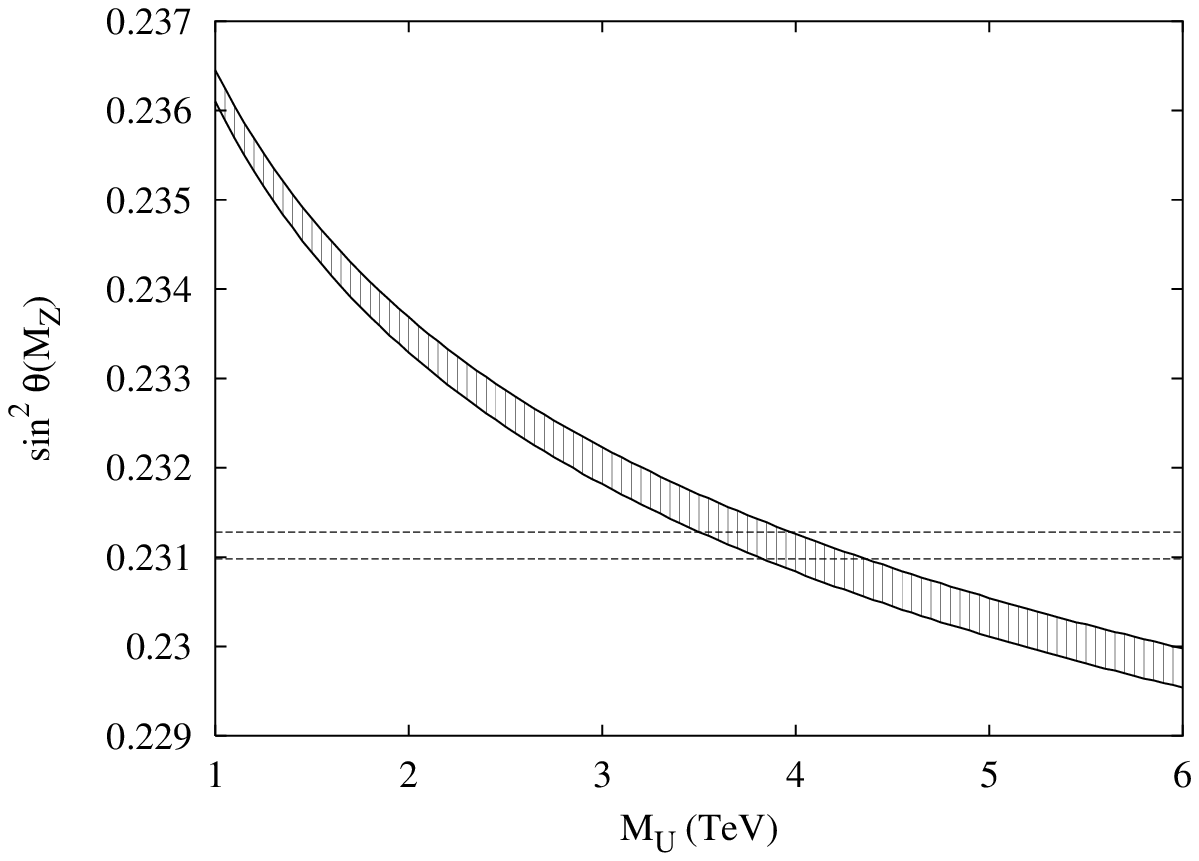}

\end{center}

\caption{}

 \end{figure}

\begin{figure}

\begin{center}

\epsfxsize=4.0in
\ \epsfbox{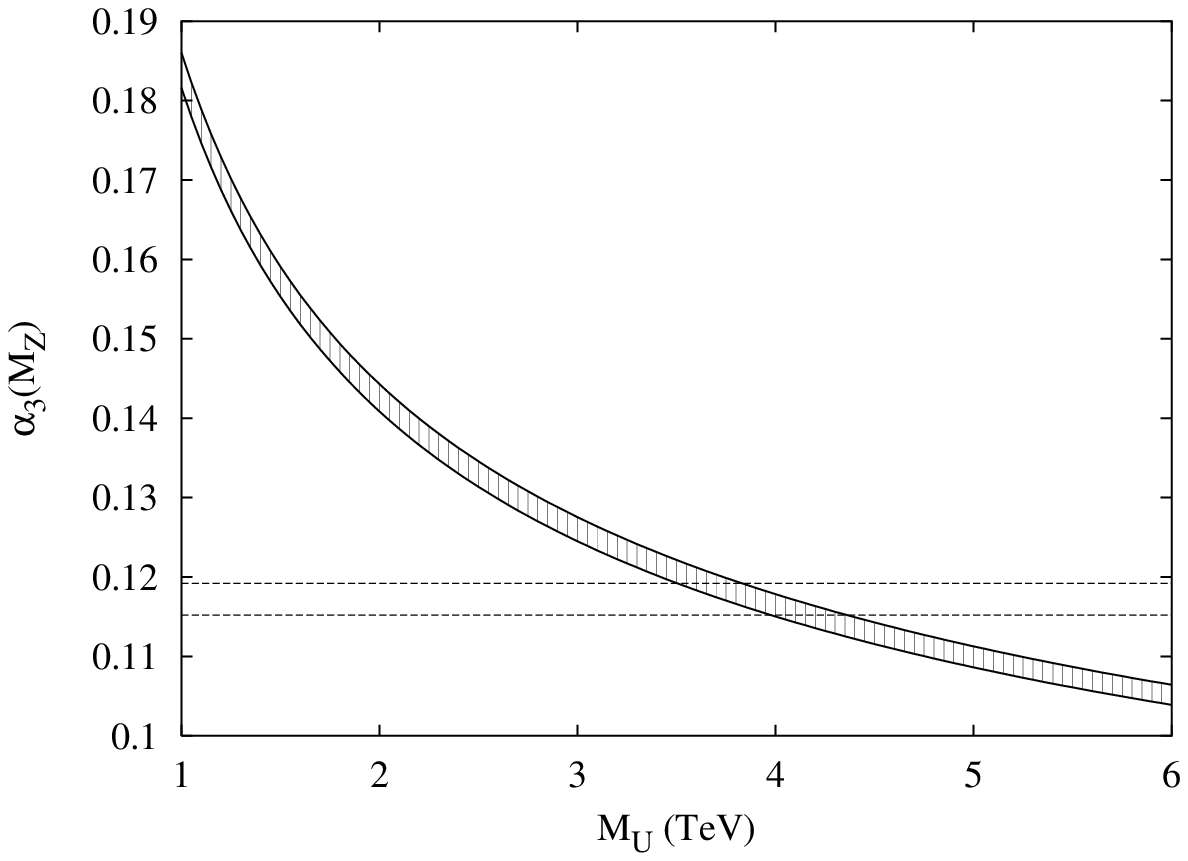}

\end{center}

\caption{}

 \end{figure}


\newpage

\begin{thebibliography}{99}
\bibitem{Maldacena}
J. Maldacena, Adv. Theor. Math. Phys. {\bf 2,} 231 (1998).
{\tt hep-th/9711200}.\\
S.S. Gubser, I.R. Klebanov and A.M. Polyakov, Phys. Lett.
{\bf B428,} 105 (1998). {\tt hep-th/9802109}.\\
E. Witten, Adv. Theor. Math. Phys. {\bf 2,} 253 (1998).
{\tt hep-th/9802150}.
\bibitem{Frampton}
P.H. Frampton, Phys. Rev. {\bf D60,} 041901 (1999). {\tt hep-th/9812117}.
\bibitem{FS}
P.H. Frampton and W.F. Shively, Phys. Lett. {\bf B454,} 49 (1999).
{\tt hep-th/9902168}.
\bibitem{FV}
P.H. Frampton and C. Vafa. {\tt hep-th/9903226.}
\bibitem{KS}
S. Kachru and E. Silverstein, Phys. Rev. Lett. {\bf 80,} 4855 (1998).
{\tt hep-th/9802183}.
\bibitem{DGG}
A. De R\'{u}jula, H. Georgi and S.L. Glashow.
{\it Fifth Workshop on Grand Unification}.
Editors: P.H. Frampton, H. Fried and K.Kang.
World Scientific (1984) page 88.
\bibitem{ADFFL}
U. Amaldi, W. De Boer, P.H. Frampton, H. F\"{u}rstenau and J.T. Liu.
Phys. Lett. {\bf B281,} 374 (1992).
\bibitem{F2}
P.H. Frampton, Phys. Rev. {\bf D60,} 085004 (1999). {\tt hep-th/9905042}.
\bibitem{mandelstam}
S. Mandelstam, Nucl. Phys. {\bf B213,} 149 (1983).
\bibitem{PS}
J.C. Pati and A. Salam,
Phys. Rev. {\bf D8,} 1240 (1973); {\it ibid} {\bf D10,} 275 (1974).
\bibitem{GG}
H. Georgi and S.L. Glashow, Phys. Rev. Lett. {\bf 32,} 438 (1974).
\bibitem{GQW}
H. Georgi, H.R. Quinn and S. Weinberg, Phys. Rev. Lett.
{\bf 33,} 451 (1974).
\bibitem{FG}
P.H. Frampton and S.L. Glashow, Phys. Lett. {\bf B131,} 340 (1983).
\bibitem{ADF}
U. Amaldi, W. de Boer and H. F\"{u}rstenau, Phys. Lett. {\bf B260,} 447 (1991).
\bibitem{Murayama}
H. Murayama and A. Pierce, Phys. Rev. {\bf D65,}
055009 (2002).
\bibitem{PP}
F. Pisano and V. Pleitez, Phys. Rev. {\bf D46,} 410 (1992).
\bibitem{PF}
P.H. Frampton,
Phys. Rev. Lett. {\bf 69,} 2889 (1992).
\bibitem{DK}
S. Dimopoulos and D. E. Kaplan, Phys. Lett. {\bf B531,} 127 (2002).
\bibitem{FMink}
P.H. Frampton and P. Minkowski, {\tt hep-th/0208024}.
\bibitem{bershadsky1}
M. Bershadsky, Z. Kakushadze and C. Vafa,
Nucl. Phys. {\bf B523,} 59 (1998).
\bibitem{bershadsky2}
M. Bershadsky and A. Johansen, Nucl. Phys. {\bf B536,} 141 (1998).
\bibitem{Witten}
E. Witten, JHEP {\bf 9812:012} (1998).
\bibitem{PDG}
Particle Data Group. {\it Review of Particle Physics.}
Phys. Rev. {\bf D66,} 010001 (2002).
\bibitem{Csaki}
C. Csaki, J. Erlich, G.D. Kribs and J. Terning,
Phys. Rev. {\bf D66,} 075008 (2002).
\bibitem{DM}
K.R. Dienes and J. March-Russell, Nucl. Phys. {\bf B479,} 113 (1996).
\bibitem{Peskin}
M. Peskin and T. Takeuchi, Phys. Rev. Lett. {\bf 65,} 964 (1990);
Phys. Rev. {\bf D46,} 381 (1992).
\bibitem{DRW}
S. Dimopoulos, S. Raby and F. Wilczek, Phys. Rev. {\bf D24,} 1681 (1981);
Phys. Lett. {\bf B112,} 133 (1982).
\bibitem{FRT}
P.H. Frampton, R.M. Rohm and T. Takahashi. {\tt hep-ph/0302XXX}.
\bibitem{DG}
S. Dimopoulos and H. Georgi, Nucl. Phys. {\bf B193,} 150 (1981).

\end{thebibliography}
\end{document}